# Co-occurrence matrices of time series applied to literary works

## Amelia Carolina Sparavigna

Department of Applied Science and Technology, Politecnico di Torino, Torino, Italy

**Abstract:** Recently, it has been proposed to analyse the literary works, plays or novels, using graphs to display the social network of their interacting characters. In this approach, the timeline of the literary work is lost, because the storyline is projected on a planar graph. However, timelines can be used to build some time series and analyse the work by means of vectors and matrices. These series can be used to describe the presence and relevance, not only of words in the text, but also of persons and places portrayed in the drama or novel. In this framework, we discuss here an approach with co-occurrence matrices plotted over time, concerning the presence of characters in the pages of a novel. These matrices are similar to those appearing in recurrence plots.

**Keywords**: Literary experiments, Time series, Co-occurrence plots, Harry Potter.

**1. Introduction**
Some approaches based on network and graph theory have been recently proposed for the analysis of literary works [1-5], adding in this manner new methods for the study of literatures. For instance, we have proposed a tool, implemented in VB.NET, able to gather information on social networks from narrative texts [6]. This tool allows finding characters and places in the narrative work by means of a list of raw words. Entering some parameters, the tool creates a social network where the nodes are the extracted characters and places, the edges their interactions. The edges are labeled by performances. An output graph is produced which can be rendered by means of Graphviz, a free open source software [5,6].
Using graphs, the timeline of the literary work is lost, being the storyline projected on the plane of the graph. However, the storyline can be used to build some time series and, by means of them, analyse the literary work with vectors and matrices. Let us remember that time series are sequences of data points, measured typically at successive points in time, points which are spaced at uniform time intervals. These series are commonly used in signal processing [7]. In researches on literary works, the time series are usually applied to the analysis of the words in the text; however, they can be used to describe the presence and relevance of characters or places in the timeline of plays and novels. As uniform time interval, we can use, for instance, the page of the book.
Time series can be analysed using frequency-domain and time-domain methods: the former includes spectral analysis, the latter auto-correlation and cross-correlation analysis. Let us consider that it is also possible to investigate the recurrence of the signal by means of recurrence plots. These plots are used to investigate natural processes (see for instance, [8]), and for the analysis of complexity and nonlinearity of physiological, cardiac and electroencephalographic signals [9-11]. In this paper, to apply the time series to the characters of literary works, we are proposing an approach with co-occurrence matrices plotted over time, which look like those of the recurrence plots. Before showing some examples, based on the first novel of the Harry Potter series written by J.K. Rowling, entitled Harry Potter and the Philosopher's Stone, let us report some researches on the use of time series in literary studies.

## 2. Time series analysis in literary studies

Today, the access to machine-readable texts is simple. We have huge electronic text archives, containing data coming from several sources and collected for various purposes. The amount of these machine-readable texts started being relevant at the end of the past century [12], but approaches to the computer-assisted content and text analysis and related methodological considerations have been developed during the sixties. Key concepts and relevant considerations were presented in a book published 1969 [12].

For the machine-readable texts, several mathematical models exist. To study the style of the literary works, as remarked in [13], it is better the models focus on features that are easily quantifiable and, for computer-aided analysis, easily identifiable by machines. Most such studies are based on frequency of occurrence of word counts, vocabulary items, or grammatical forms [13]. Besides the simple analysis of style, in the Reference 13, it is discussed the use of the autoregressive integrated moving average (ARIMA) models, which are traditionally used to describe correlated time series data. In particular, in statistics and econometrics, the ARIMA models are used to predict future points in the series (forecasting) [14]. The author of [13] developed ARIMA models for different works by the same author and works by different authors, based on the correlations of the number of words in each sentence.

Besides being fundamental for the analysis of a large number of works, the use of computation avoids the problem of bias created by unpredictable autocorrelation. As told in [15], autocorrelation can inflate or deflate the differences existing among the different parts of the texts under analysis, biasing the conclusions.

Time series have been also used to quantify the rhythm of poems and novels [16]: in this reference, the various rhythms are characterized by means of linear and nonlinear variability measures. These measures are applied to a time series representation of the literary work whit the dependent variable being the number of characters in each word of the literary text [16]. Variability measures are the fluctuation index and the approximate entropy, which estimate the amount of regularity and the unpredictability of fluctuations [17]. In [16], some empirical evidences are showing that word length is an appropriate representation of rhythm, and demonstrate that the fluctuation index and approximate entropy are effective quantifiers of rhythm.

## 3. Matrices and characters

Besides the series of words, we can have series from the presence of characters and places in the timeline of novels. In this case, a possible choice for a uniform time interval to use in the series can be the page of the book. This approach is useful to integrate the information coming from the network of characters and their interactions. As previously told, the network of characters, displayed by a planar graph, can be used to visualize the overall social network of the play or novel. This is a representation where the time is lost. To solve this problem, we could create a graph for each act or chapter of the literary work considered, but this could require, in some cases, a lot of images to show them. An alternative and more compact proposal is the use of occurrence and co-occurrence matrices plotted over time. These matrices can show the dynamics of the characters.

Some co-occurrence matrices for literary works are shown in [18], where the author provides a matrix diagram visualizing character co-occurrences in Victor Hugo's Les Misérables. There, each colored cell represents two characters that appeared in the same chapter. However, the matrices are built on the name of characters, not on time. To have a dynamics we need time, and, if possible, co-occurrence between characters and places too.

In the following, we will show the use of co-occurrence matrices of characters built on time.

## 4. Harry Potter's time series

To illustrate time series and their co-occurrence matrices, let us consider some examples based on "Harry Potter and the Philosopher's Stone" (or Sorcerer's Stone). The story is supposed to be well known to the readers: in any case, the plot and a partial list of characters are given in [19]. The book is J.K. Rowling's debut novel, published on 26 June 1997.

In [4], we have discussed the network of the characters; details of nodes and edges are given in the Appendix of the reference. The central character is Harry Potter (a hub, or pivot of the novel). In the network, Harry Potter has the highest degree, quite larger than those of Ron, Hagrid and Hermione. He is the center of two clusters, one is concerning his life with Dursleys, in the world of the non-magical people, and the other is concerning his life at the Hogwarts School.

In the book, Harry, the young wizard, is present in almost every page and therefore a time series of him is not interesting. It is more relevant to prepare the time series for the other main characters, such as his friends Ron Weasley, Hermione Granger, Hagrid the gamekeeper and Professor Dumbledore, and of his "enemies" Dudley Dursley and Draco Malfoy. Let us suppose a hypothetic text containing the novel in 200 pages. In the Figure 1, we can see the time series representing the occurrence in each of the 200 pages of the characters. In fact, it is an occurrence function assuming two values, zero and one, given in the domain of the number of pages. The time series looks like a barcode. If the name of the characters appears in the page once or more times, the function has a value equal to one, otherwise, the value of the function is zero. Given a page, if the function is zero we have a white bar, if the value is one we have a black bar. The close friends Ron and Hermione have a code dense of black bars in the second part of the novel, which occupies the two thirds of the novel. This part tells about Harry at the Hogwarts School. In the first part, we find the character of Dudley. We see Hagrid in all the book; in fact, he is the person who takes care of Harry, after he leaves the Dursleys and before his arrival at the Hogwarts School. And of course, being the gamekeeper at this School he is also present in the second part of the novel.

## 5. Harry Potter's occurrence and co-occurrence matrices

As told before, time series are also investigated using the recurrence plots (RPs). RPs are useful in the analysis of the recurrence of a signal, in particular of natural and physiological processes [8-11]. In statistics, a recurrence plot is defined as a plot showing the times at which a phase space trajectory visits roughly the same area in the phase space. Extensions of recurrence plots are the cross recurrence plots, which are plots considering the phase space trajectories of two different systems in the same phase space [8,20]. RPs and CRPs are square plots where both axes are time axes.

Recurrence plots are useful for data files that contain a large number of data points, such as those given in Reference 8, to reveal complex spatial and temporal details of the time series. In the case of time series extracted from a literary work, RPs can be useful in the analysis of rhythms in the text [16]. In the case of time series of the Figure 1, the use of RPs is not appropriate, however we can model on them the occurrence matrices. These matrices are shown in the Figure 2: we see square plots, where both axes are time axes, the time being represented by the number of the page. The black dots represent the occurrence of the character in the page of the novel. Of course, the occurrence matrices provide the same information of the barcodes.

Using different colours, we can represent on the same plot the co-occurrence of several characters. In the Figure 3, for instance, we have the co-occurrence plots over time of Hagrid with Dumbledore, Dudley, Ron, Hermione and Draco. The pink colour represents the occurrences of Hagrid, whereas the light blue the presence of the other character. The black dots represent the co-occurrence of both characters in the same page; therefore, these plots are showing interactions between characters and when these interactions are occurring. In the same figure, we see the co-occurrence matrices of Ron with Dumbledore, Dudley, Hagrid, Hermione and Draco: in this case, the pink colour represents the occurrence of Ron and the light blue that of the other character.
Co-occurrence plots can be used for more than two characters. In the Figure 4 we are showing an example where there are three characters: Hagrid, Ron and Hermione. The black dots are showing the presence of the three in the same page of the novel.

**6. Conclusion**
Here, in applying the time series to the characters of literary works, we are proposing an approach based on occurrence and co-occurrence matrices plotted over time. These matrices are modelled on recurrence and cross-recurrence plots. We have shown some examples, using the novel "Harry Potter and the Philosopher's Stone", with the aim of illustrating a method suitable to display the interactions among characters during the novel, and therefore integrating the graphs of the social networks with information concerning time. Some further studies will be devoted to the use of recurrence plots and cross-recurrence plots of characters and places in literary works.

**References**
1. F. Moretti, Network theory, plot analysis, A Stanford Lit Lab Pamphlet, 2011, http://litlab.stanford.edu/LiteraryLabPamphlet2A.Text.pdf
2. J. Stiller, D. Nettle and R.I.M. Dunbar, The small world of Shakespeare's plays, Human Nature, Volume 14, Issue 4, 2003, pages 397-408.
3. R. Alberich, J. Miro-Julia and F. Rossello, Marvel Universe looks almost like a real social network, http://arxiv.org/abs/cond-mat/0202174v1.
4. A.C. Sparavigna, On Social Networks in Plays and Novels, The International Journal of Sciences, 2013, Volume 2, Issue 10, Pages: 20-25.
5. A.C. Sparavigna and R. Marazzato, Graph Visualization Software for Networks of Characters in Plays, The International Journal of Sciences, 2014, Volume 3, Issue 2, Pages: 69-79.
6. R. Marazzato and A.C. Sparavigna, Extracting Networks of Characters and Places from Written Works with CHAPLIN, arXiv:1402.4259 [cs.CY], 2014.
7. G. Box and G. Jenkins, Time Series Analysis Forecasting and Control, Holden-Day, San Francisco, 1976.
8. A.C. Sparavigna, Recurrence Plots of Sunspots, Solar Flux and Irradiance, arXiv:0804.1941 [physics.pop-ph], 2008.
9. C.L. Webber Jr and J.P. Zbilut, Dynamical Assessment of Physiological Systems and States Using Recurrence Plot Strategies, Journal of Applied Physiology, 1994, Volume 76, Issue 2, Pages 965-973.
10. J.P. Zbilut, N. Thomasson, and C.L. Webber, Recurrence Quantification Analysis as a Tool for Nonlinear Exploration of Nonstationary Cardiac Signals, Medical Engineering & Physics, 2002, Volume 24, Issue 1, Pages: 53-60.
11. Gaoxiang Ouyang, Li Xiaoli, Dang Chuangyin and D.A. Richards, Using Recurrence Plot for Determinism Analysis of EEG Recordings in Genetic Absence Epilepsy Rats, Clinical Neurophysiology, 2008, Volume 119, Issue 8, Pages 1747-1755.
12. Melina Alexa, Computer-assisted text analysis methodology in the social sciences, ZUMA-Arbeitbericht 97/07, October 1997.
13. R. Oppenheim, The Mathematical Analysis of Style: A Correlation-based Approach, Computers and the Humanities, 1988, Volume 22, Issue 4, pp 241-252
14. Vv. Aa., Wikipedia, /wiki/Autoregressive_integrated_moving_average

15. R. Hogenraad, D.P. McKenzie and C. Martindale, The Enemy within: Autocorrelation Bias in Content Analysis of Narratives, Computers and the Humanities, 1996/1997, Volume 30, Issue 6, pp. 433-439
16. R. Pitz, Quantifying Degrees of Randomness in Word Rhythms of Literary Works, ProQuest, UMI Dissertation Publishing, 2008.
17. Vv. Aa., Wikipedia, wiki/Approximate_entropy
18. M. Bostock, Les Misérables Co-occurrence, at http://bost.ocks.org/mike/miserables/
19. http://en.wikipedia.org/wiki/Harry_Potter_and_the_Philosopher's_Stone
20. N. Marwan, M.C. Romano, M. Thiel and J. Kurths, Recurrence Plots for the Analysis of Complex Systems, Physics Reports, 2007, Volume 438, Issue 5-6, Pages: 237-329.

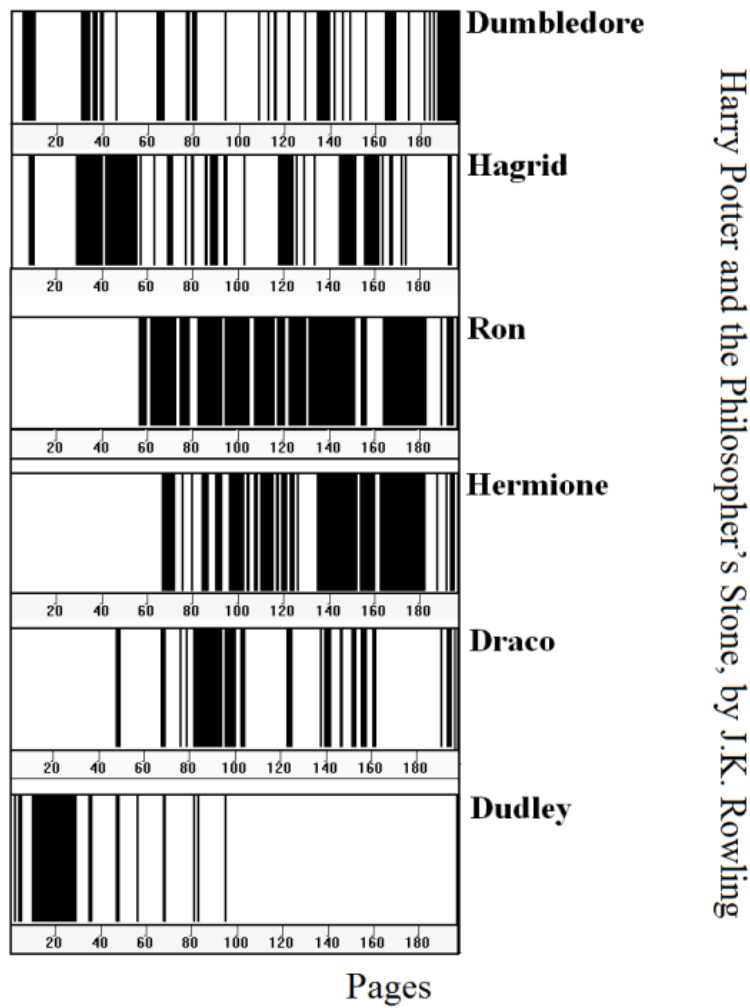

**Figure 1** - Time series of Harry's friends Ron Weasley, Hermione Granger, Hagrid the gamekeeper and Professor Dumbledore, and also of his "enemies" Dudley Dursley and Draco Malfoy. We are supposing a hypothetic text containing the novel in 200 pages. The functions represent the occurrence in a page of the characters in a barcode-like image. If the name of the characters appears in the page once or more times, the function has a value equal to one, otherwise, the value of the function is zero. Given a page, if the function is zero we have a white bar, if the value is one we have a black bar.

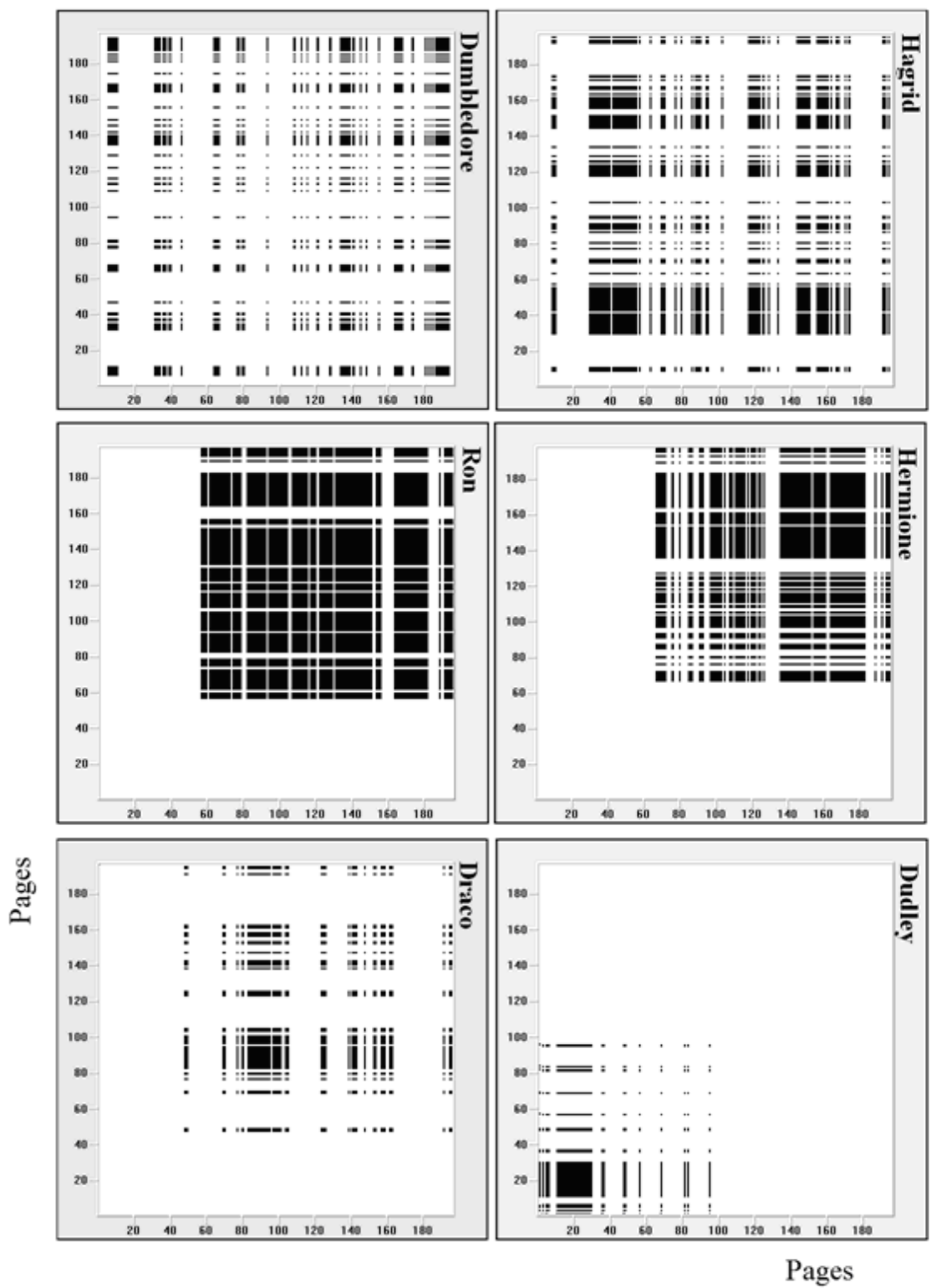

**Figure 2** - Recurrence plots are useful to reveal complex spatial and temporal details of the time series. Having these plots as models, we can create some occurrence matrices for time series shown in the Figure 1. Here we see square plots, where both axes are time axes; the number of the page represents time. The black dots represent the occurrence of the character in the page of the novel. Of course, these matrices provide the same information of the barcodes.

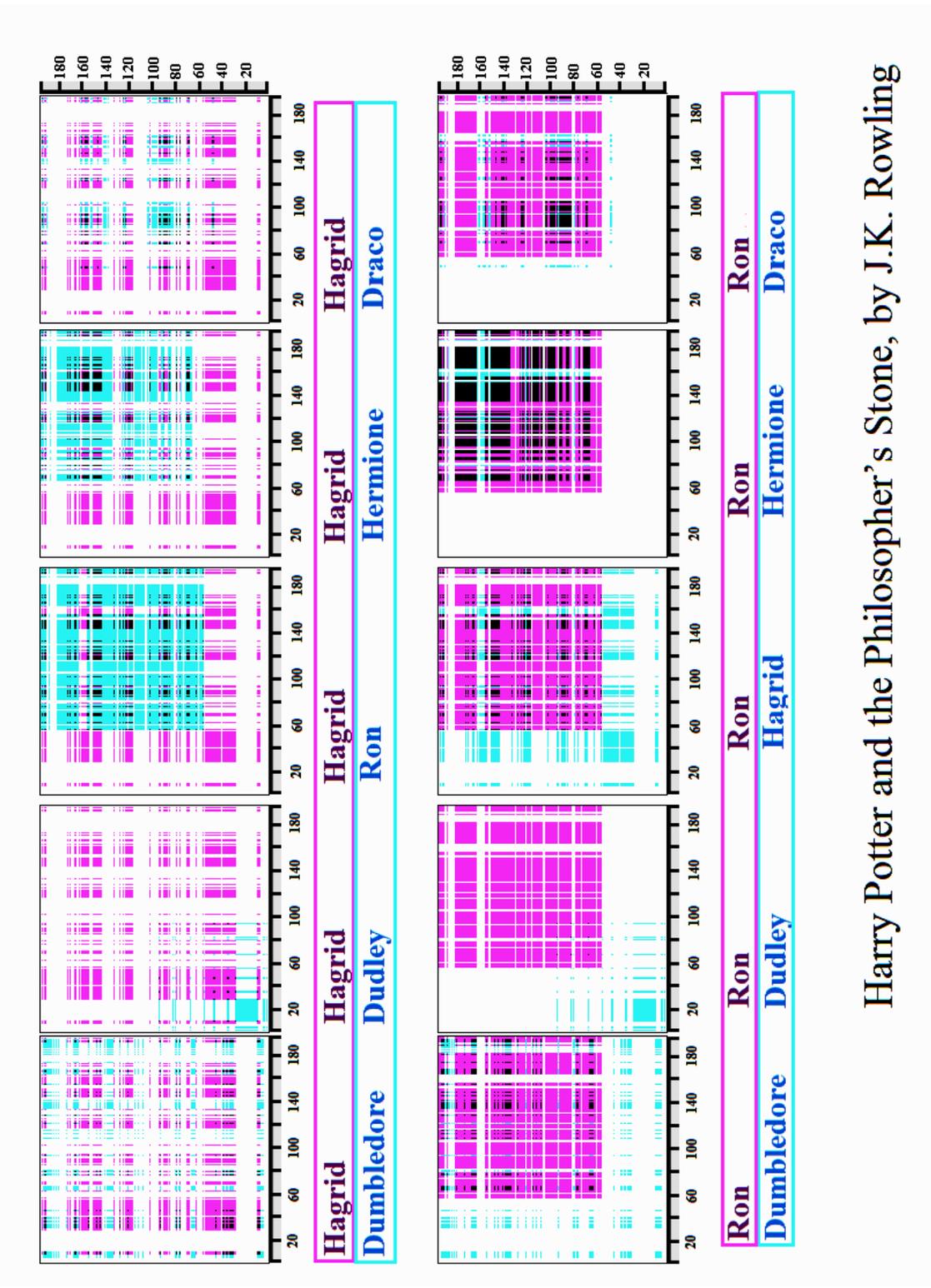

**Figure 3** - Using different colours, we can represent on the same plot the co-occurrence of several characters. Here we have the co-occurrence plots over time of Hagrid with Dumbledore, Dudley, Ron, Hermione and Draco. The pink colour represents Hagrid's occurrences, whereas the light blue that of the other character. The black dots represent the co-occurrence of both characters in the same page. We can see also the co-occurrence matrices of Ron with Dumbledore, Dudley, Hagrid, Hermione and Draco: in this case, the pink colour represents the occurrence of Ron and the light blue that of the other character.

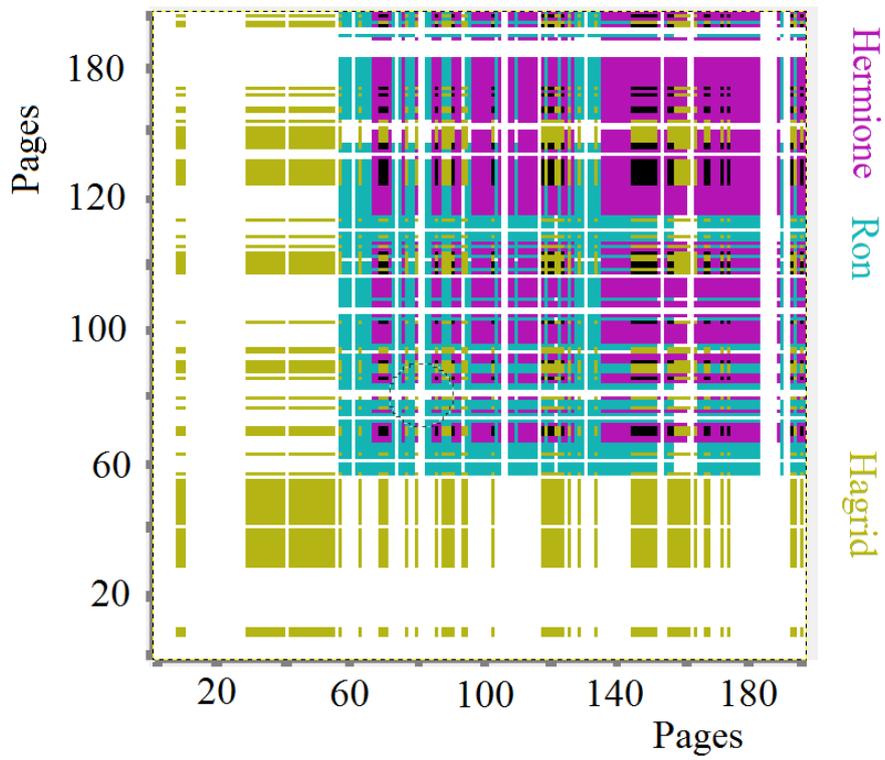

**Figure 4** – Co-occurrence plot for three characters: Hagrid, Ron and Hermione. The black dots represent the presence of them in the same page.